\documentclass{aastex}
\usepackage{times}
\def\mincir{\raise -2.truept\hbox{\rlap{\hbox{$\sim$}}\raise5.truept \hbox{$<$}\ }}
\def\mincireq{\hbox{\raise0.5ex\hbox{$<\lower1.06ex\hbox{$\kern-1.07em{\sim}$}$}}}
\def\magcir{\raise-2.truept\hbox{\rlap{\hbox{$\sim$}}\raise5.truept \hbox{$>$}\ }}
\def\gr{\kern 2pt\hbox{}^\circ{\kern -2pt K}} 

\def\_{\thinspace}

\shorttitle{SED vs. activity in blazars }
\shortauthors{Mankuzhiyil et al.}

\begin{document}

\title{THE ENVIRONMENT AND DISTRIBUTION\\
OF EMITTING ELECTRONS\\
AS A FUNCTION OF SOURCE ACTIVITY\\
IN MARKARIAN 421}

\author{Nijil\,Mankuzhiyil\altaffilmark{1}}
\affil{Dipartimento di Fisica, Universit\`{a} di Udine, via delle Scienze 208, I-33100 Udine (UD), ITALY}

\author{Stefano\,Ansoldi\altaffilmark{1,2}}
\affil{International Center for Relativistic Astrophysics (ICRA)}

\author{Massimo\,Persic\altaffilmark{1}}
\affil{INAF-Trieste, via G.\,B.\,Tiepolo 11, I-34143 Trieste (TS), ITALY}

\author{Fabrizio\,Tavecchio}
\affil{INAF-Brera, via E.\,Bianchi 46, I-23807 Merate (LC), ITALY}

\altaffiltext{1}{and INFN Sezione di Trieste}
\altaffiltext{2}{and Dipartimento di Matematica e Informatica, Universit\`{a} di Udine,
via delle Scienze 206, I-33100 Udine (UD), ITALY}


\begin{abstract}
For the high-frequency peaked BL\,Lac object Mrk\,421 we study the variation of
the spectral energy distribution (SED) as a function of source activity, from
quiescent to active. We use a fully automatized $\chi^2$-minimization procedure,
instead of the ``eyeball'' procedure more commonly used in the literature, to model
nine SED datasets with a one-zone Synchrotron-Self-Compton (SSC) model and examine
how the model parameters vary with source activity. The latter issue can finally
be addressed now, because simultaneous broad-band SEDs (spanning from optical to
VHE photon energies) have finally become available. Our results suggest that in
Mrk\,421 the magnetic field ($B$) decreases with source activity, whereas the
electron spectrum's break energy ($\gamma_{\mathrm{br}}$) and the {Doppler} factor
($\delta$) increase -- the other SSC parameters turn out to be uncorrelated with
source activity. In the SSC framework these results are interpreted in a picture
where the synchrotron power and peak frequency remain constant with varying source
activity, through a combination of decreasing magnetic field and increasing number
density of $\gamma \leq \gamma_{\mathrm{br}}$ electrons: since this leads to an increased
electron-photon scattering efficiency, the resulting Compton power increases, and
so does the total ($=$ synchrotron plus Compton) emission.
\end{abstract}


\section{Introduction}

It's commonly thought that the fueling of supermassive black holes,
hosted in the cores of most galaxies, by surrounding matter produces
the spectacular activity observed in AGNs. In some cases ($\mincir$10\%)
powerful collimated jets shoot out in opposite directions at relativistic
speeds. The origin of such jets is one of the fundamental open problems
in astrophysics.

If a relativistic jet is viewed at small angle to its axis, the
observed emission is amplified by relativistic beaming (Doppler
boosting and aberration) allowing deep insight into the physical
conditions and emission processes of relativistic jets. Sources
whose boosted jet emission dominates the observed emission (blazars%
\footnote{
    Within the blazar class, extreme objects,
    lacking even signatures of thermal processes
    usually associated with emission lines, are
    defined as BL\,Lac objects.}%
) represent a minority among AGN, but are the dominant extragalactic
source class in $\gamma$-rays. Since the jet emission overwhelms all
other emission from the source, blazars are key sources for studying
the physics of relativistic extragalactic jets.

The jets' origin and nature are still unclear. However, it is widely believed
that jets are low-entropy (kinetic/electromagnetic) flows that dissipate some
of their energy in (moving) regions associated with internal or external shocks.
This highly complex physics is approximated, for the purpose of modelling the
observed emission, with one or more relativistically moving homogeneous plasma
regions (blobs), where radiation is emitted by a non-thermal population of particles
\citep[e.g.][]{Mar+92}. The high energy emission, with its extremely fast and
correlated multifrequency variability, indicates that often one single region
dominates the emission.

\begin{figure*}
\begin{center}
\includegraphics[width=14cm]{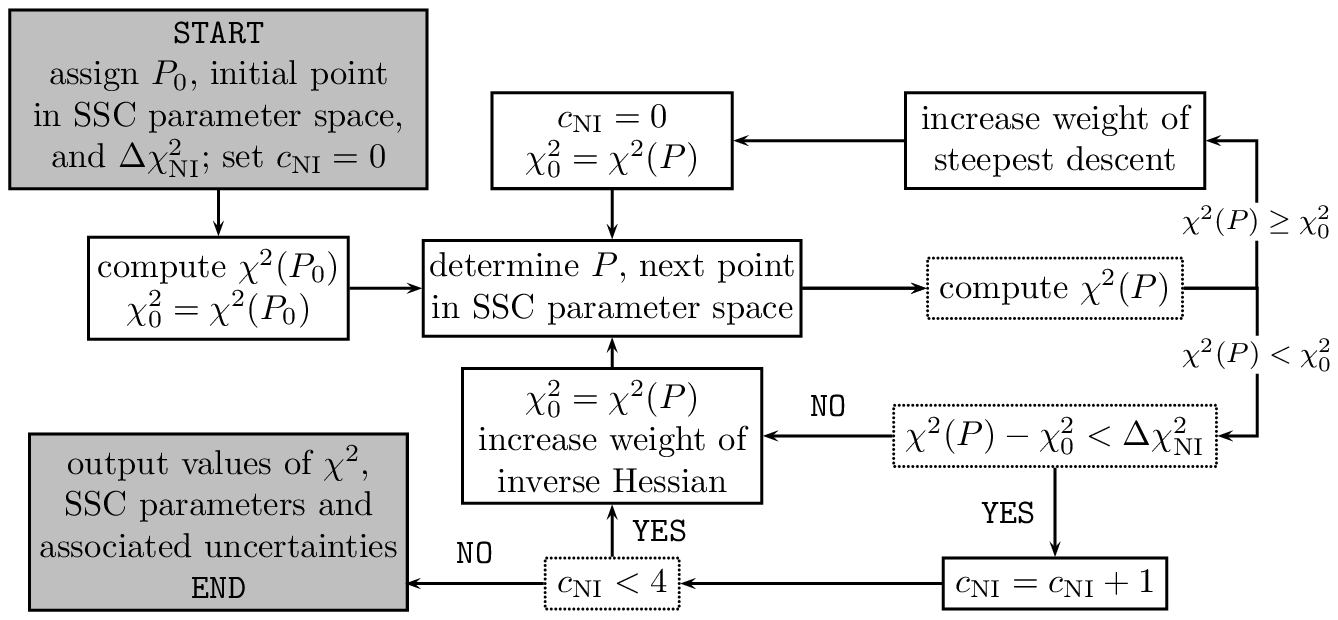}
\caption{\label{fig:flowchart}Flow chart of the minimization code.}
\end{center}
\end{figure*}

The jet's broad-band (from radio to $\gamma$-ray frequencies) spectral energy
distribution (SED) is a non thermal continuum featuring two broad humps that
peak at IR/X-ray and GeV/TeV frequencies
and show correlated luminosity and spectral changes. This emission is commonly
interpreted within a Synchrotron-Self-Compton (SSC) model where
the synchrotron and Compton peaks are produced by one same time-varying
population of particles moving in a magnetic field \citep[e.g.,][hencefort T98]{Tav+98}.
The SSC model is perfectly adequate to explain the SEDs of BL\,Lac
sources \citep[e.g.,][]{Tav+10}.

One important issue that should be addressed, but has not yet been so far
because of the lack of simultaneous broad-band SEDs, is how the emission changes as a
function of the source's global level of activity. In particular, given an emission model
that fits the data, it is should be examined what model parameters are correlated with
source activity.
In order to investigate SEDs at different levels of activity we choose a
high-frequency-peaked BL\,Lac (HBL) object, i.e. a blazar
(i) whose relativistic jet points directly toward the observer, so owing to
relativistic boosting its SSC emission dominates the source;
(ii) whose Compton peak ($\magcir$100\,GeV) can be detected by Cherenkov
telescopes; and
(iii) whose GeV spectrum can be described as a simple power law \citep[unlikely
other types of BL\,Lacs, see][]{Abdo+09}. In addition, such HBL source
must have several simultaneous SED datasets available.
Mrk\,421 meets these requirements. In this paper we study the variation of its
SED with source activity, from quiescent to active.

Another requirement for this kind of study concerns the modeling procedure. We use a
full-fledged $\chi^2$-minimization procedure instead of the "eyeball" fits
more commonly used in the literature. While the latter at most prove the
existence of a {\it good} solution, by finely exploring the parameter space
our procedure finds the {\it best} solution and also proves such solution
to be {\it unique}.

In this paper we investigate the SED of Mrk\,421 in nine different source states.
To this aim we fit a one-zone SSC emission model (described in Sect.2), using a
fully automatized $\chi^2$-minimization procedure (Sect.3), to the datasets described
in Sect.4. The results are presented and discussed in Sect.5.

\section{BL\,Lac SSC emission}

To describe the HBL broad-band emission, we use the one-zone SSC model of T98.
This has been shown to adequately describe
broad-band SEDs of most high-frequency-peaked BL\,Lac objects \citep[e.g.,][]{Tav+10}
and, for a given source, both its ground and excited states \citep{Tav+01, Taglia+08}.
The main support for the one-zone model is that in most such sources the temporal
variability is clearly dominated by one characteristic timescale, which implies one
dominant characteristic size of the emitting region \citep[e.g.,][]{Ander+09}.
Moreover, one of the most convincing evidence favoring the SSC model is the strict
correlation that usually holds between the X-ray and VHE$\, \gamma$-ray variability
\citep[e.g.,][]{Foss+08}: since in the SSC model the emission in the two bands is
produced by the same electrons (via synchrotron and SSC mechanism, respectively), a
strict correlation is expected%
\footnote{The rarely occurring ``orphan'' TeV flares, that are not accompanied by
    variations in the X-ray band, may arise from small, low-$B$,
    high-density plasma blobs \citep{Kraw+04}.}.

{
In our work, for simplicity we used a one-zone SSC model, assuming that
the entire SED is produced within a single homogeneous region of the jet.
As already noted, this class of models is generally adequate to reproduce
HBL SEDs. However, one-zone models also face some problems in explaining
some specific features of TeV blazar emission. In particular, while very
large Doppler factors are often required in one-zone model, radio VLBI
observations hardly detect superluminal motion at parsec scale (e.g.,
\citep{Piner+10, Gir+06}). This led \citep{GeorKaz03, Ghis+05} to propose
the existence of a structured, inhomogeneous and decelerating emitting jet.
Inhomogeneous (two-zone) models (e.g., \citep{GhisTav08}) have been also
invoked to explain the ultra-rapid variability occasionally observed in TeV
blazars (e.g., \citep{Ahar+07, Albert+07}.
}

The emission zone is supposed to be spherical with radius $R$, in relativistic motion with
bulk Lorentz factor $\Gamma$ at an angle $\theta$ with respect to the line of sight to the
observer, so that special relativistic effects are cumulatively described by the relativistic
Doppler factor, $\delta=[\Gamma(1-\beta\,{\rm cos}\,\theta)]^{-1}$. Relativistic electrons
with density $n_{\rm e}$ and a tangled magnetic field with intensity $B$ homogeneously fill
the region.

The relativistic electrons' spectrum is described by a smoothed broken power-law function of
the electron Lorentz factor $\gamma$, with limits $\gamma_1$ and $\gamma_2$, break at
$\gamma_{\rm br}$ and low- and high-energy slopes $n_{1}$ and $n_{2}$. This purely
phenomenological choice is motivated by the observed shape of the humps in the SEDs, well
represented by two smoothly joined power laws for the electron distribution.
Two ingredients are important in shaping the VHE\,$\gamma$-ray part of the spectrum: (i)
using the Thomson and Klein-Nishina cross sections when, respectively, $\gamma h \nu \leq
m_{\rm e}c^2$ and $\gamma h \nu > m_{\rm e}c^2$ (with $\nu$ the frequency of the 'seed'
photon), in building the model (T98); and (ii) the correction of $\gtrsim 50\,$GeV data for
absorption by the Extragalactic Background Light (EBL), as a function of photon energy and
source distance \citep[e.g.,][and references therein]{Mank+10}: for this purpose we use the
popular \cite{Fran+08} EBL model.

The one-zone SSC model can be fully constrained by using simultaneous multifrequency observations
(e.g., T98). Of the nine free parameters of the model, six specify the
electron energy distribution ($n_{\rm e}$, $\gamma_1$, $\gamma_{\rm br}$, $\gamma_2$, $n_1$,
$n_2$), and 3 describe the global properties of the emitting region ($B$, $R$, $\delta$).
Ideally, from observations one can derive six observational quantities that are uniquely linked
to model parameters: the slopes, $\alpha_{1,2}$, of the synchrotron bump at photon energies
below and above the UV/X-ray peak are uniquely connected to $n_{1,2}$; the synchrotron and
SSC peak frequencies, $\nu_{\rm s,C}$, and luminosities, $L_{\rm s,C}$, are linked with $B$,
$n_{\rm e}$, $\delta$, $\gamma_{\rm br}$; finally, the minimum variability timescale $t_{\rm var}$
provides an estimate of the source size through $R \mincir c t_{\rm var} \delta /(1+z)$.

To illustrate how important it is to sample the SED around {\it both} peaks, let us consider a
standard situation before Cherenkov telescopes came online, when we would have had only knowledge
of the UV/X-ray (synchrotron) peak. Clearly this would have given us information on the shape of
the electron distribution but would have left all other parameters unconstrained: in particular
the degeneracy between $B$ and $n _{\mathrm{e}}$ -- inherent in the synchrotron emissivity --
could not be lifted without the additional knowledge of the HE/VHE\,$\gamma$-ray (Compton) peak.

Therefore, only knowledge of observational quantities related to both SED humps enables
determination of all SSC model parameters.

\section{$\chi^2$ minimization}

In this section we discuss the code that we have used to obtain an estimation of the
characteristic parameters of the SSC model. As we recalled in the previous section
the SSC model that we are assuming is characterized by nine free parameters,
$n_{\rm e}$, $\gamma_1$, $\gamma_{\rm br}$, $\gamma_2$, $n_1$, $n_2$, $B$, $R$, $\delta$.
However, in this study we set $\gamma_{\rm 1} = 1$, which is a widespread assumption in the literature,
so reducing the number of free parameters to eight.

The determination of the eight free parameters has been performed by finding
their best values and uncertainties from a $\chi ^{2}$ minimization
in which multi-frequency experimental points have been fitted to the SSC model
described in T98. Minimization has been performed using the
Levenberg--Marquardt method \citep[see][]{Press+94}, which is an efficient standard for
non-linear least-squares minimization that smoothly interpolates between two different
minimization approaches, namely the inverse Hessian method and the steepest
descent method.

{
The algorithm starts by making an educated guess for a starting point $P _{0}$ in parameter
space with which the minimization loop is entered; at the same time a (small) constant is
defined, $\Delta \chi^2_{\rm{NI}}$ (where NI stands for 'Negligible Improvement'), which
represents an increment in $\chi^2$ small enough that the minimization step can be considered
to have achieved no significative improvement in moving toward the minimum $\chi^2$: this will
be used as a first criterion for the exit condition from the minimization loop. Indeed, to be
more confident that the minimization loop has reached a $\chi^2$ value close enough to the
absolute minimum, the above condition has to be satisfied four times in a row: the number of
consecutive times the condition has been satisfied is conveniently stored into the integer
variable $c _{\rm{NI}}$ -- which is, thus, set to zero at startup. From the chosen value of
$P_0$ we can compute the associated $\chi^2_0 = \chi^2(P_0)$ and enter the minimization loop.
The Levenberg-Marquardt method will determine the next point in parameter space, $P$, where
$\chi^2$ will be evaluated. If $\chi^2(P) > \chi^2_0$ the weight of the steepest-descent method
in the minimization procedure is increased, the variable $c_{\mathrm{NI}}$ is set to zero, and
we can proceed to the next minimization step. If, instead, $\chi^2(P) \leq \chi^2_0$ we further
check if the decrease in $\chi^2$ is smaller than or equal to $\Delta \chi^2_{\rm{NI}}$. If this is the
case, the negligible-improvement counter $c _{\rm{NI}}$ is increased by one: if the resulting
value is $c _{\rm{NI}} \geq 4$, we think we have a good enough approximation of the absolute
minimum -- and the algorithm ends. If, instead, in the latter test $c _{\rm{NI}} < 4$, or if in
the preceding test $\Delta \chi^2 > \Delta \chi^2_{\rm{NI}}$, then we increase the weight of the
inverse Hessian method in the minimization procedure, we set $\chi^2_0$ equal to the lower value
we have just found, and we continue the minimization loop.
}
For completeness {and illustration} , we briefly present the flow-chart of the algorithm in
Fig.~\ref{fig:flowchart}.

A crucial point in our implementation is that from T98 we can
only obtain a numerical approximation to the SSC spectrum, in the form of a sampled
SED. On the other hand, at each step of the loop (see Fig.~\ref{fig:flowchart}) the calculation of
$\chi ^{2}$ requires the evaluation of the SED for all the observed frequencies. Usually
the model function is known analytically, so these evaluations are a straightforward
algebraic process. In our case, instead, we know the model function only through a
numerical sample and it is unlikely that an observed point will be one of the sampled
points coming from the implementation of the T98 model. Nevertheless it will in general
fall between two sampled points, which allows us to use interpolation%
\footnote{
    The sampling of the SED function derived from T98 is dense enough,
    so that, with respect to other uncertainties, the one coming from
    this interpolation is negligible.}
to approximate the value of the SED.

At the same time, the Levenberg--Marquardt method requires the calculation of the partial derivatives
of $\chi ^{2}$ with respect to the eight fitted SSC parameters. Contrary to the usual case, in which from the
knowledge of the model function all derivatives can be obtained analytically, in our case they have also
been obtained numerically by evaluating the incremental ratio of the $\chi ^{2}$ with respect to a sufficiently
small, dynamically adjusted increment of each parameter. This method could have introduced a potential
inefficiency in the computation, due to the recurrent need to evaluate the SED at many, slightly
different points in parameter space, this being the most demanding operation in terms of CPU time.
For this reason we set up an algorithm to minimize the number of calls to T98 across different
iterations.

\begin{figure*}
\begin{center}
\includegraphics[width=15.8cm]{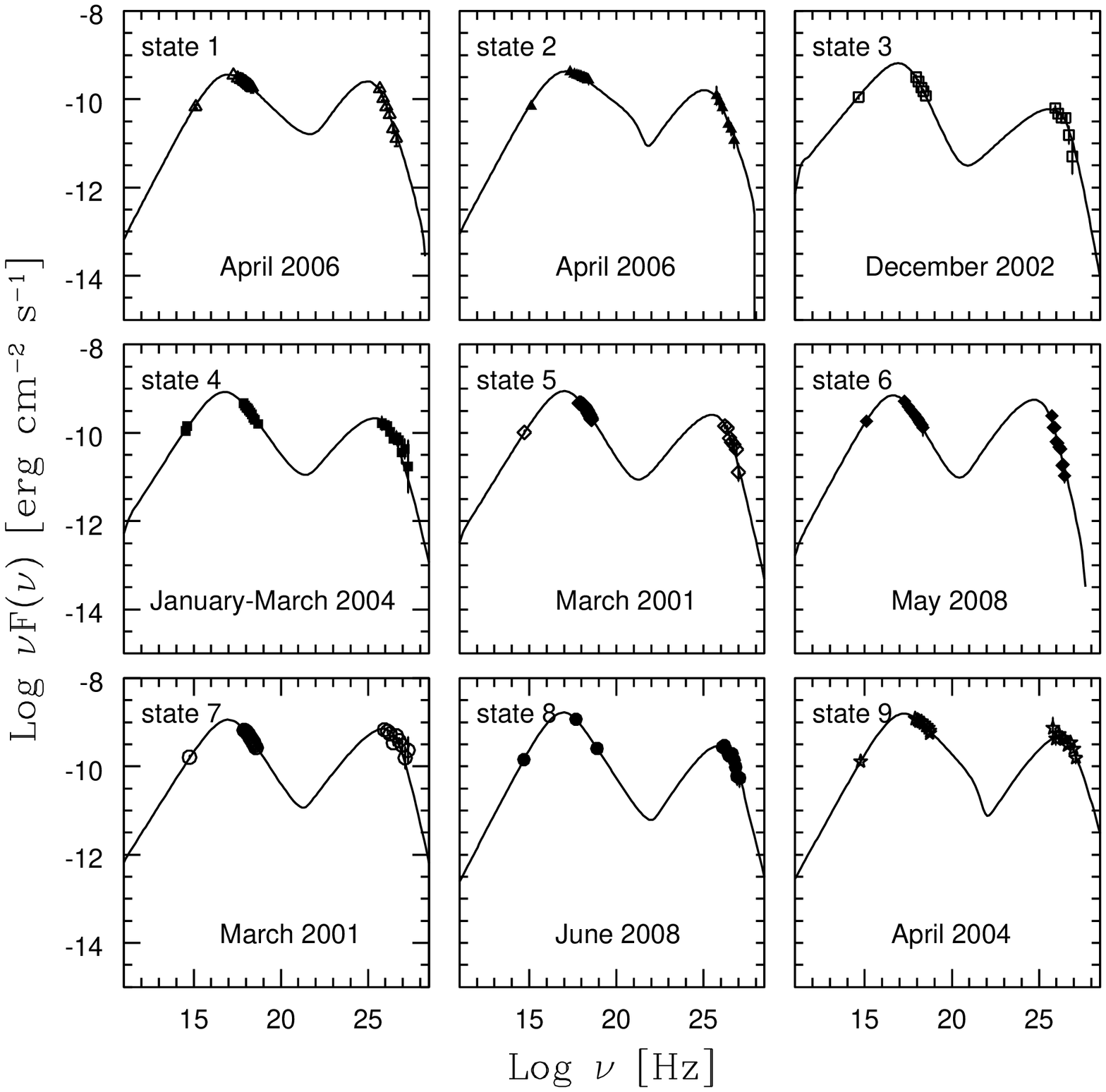}
\caption{\label{fig:bestfit00}Best-fit one-zone SSC models for nine data sets referring to different
emission levels of the HBL source Mrk\,421. Source states are ordered by increasing
model luminosity and data have been obtained as follows: state 1 \citep{flaring_state126},
state 2 \citep{flaring_state126},
state 3 \citep{flaring_state3},
state 4 \citep{flaring_state49},
state 5 \citep{Foss+08},
state 6 \citep{flaring_state126},
state 7 \citep{Foss+08},
state 8 \citep{flaring_state8},
state 9 \citep{flaring_state49}.}
\end{center}
\end{figure*}

\section{Datasets}

From the literature we then select nine SED datasets corresponding to different emission states
(low to high) of the HBL source Mrk\,421.

{
State\,1 and state\,2 \citep{flaring_state126} multi-wavelength campaigns were triggered by a major outburst in
April 2006 that was detected by the {\it Whipple} 10\,m telescope. A prompt campaign was not possible because of
visibility constraints on XMM-{\it Newton}. So simultaneous multi-wavelength observations took place during the
decaying phase of the burst. The optical/UV and X-ray observations were carried out using XMM-{\it Newton}'s optical
monitor (OM) and EPIC-pn detector, respectively. The MAGIC and {\it Whipple} telescopes were used for the
VHE\,$\gamma$-ray observations. State\,1 strictly simultaneous observations lasted $\sim$4\,hrs for state\,1, and
more than 3\,hrs for state\,2.

State\,3 \citep{flaring_state3} reports the multi-wavelength observations during December 2002 and January 2003.
The campaign was initiated by X-tay and VHE flares detected by the All Sky Monitor (ASM) of the {\it Rossi} X-ray
Timing Explorer ({\it R}XTE) and the 10\,m {\it Whipple} telescope. {\it Whipple} and HEGRA-CT1 were used for the
VHE observations during the campaign. Even though {\it Whipple} observed the source from Dec.4, 2002 to Jan.15,
2003, and HEGRA-CT1 on Nov.3--Dec.12, 2002, only the data taken during nights with simultaneous X-ray observations
are used in this paper to construct the SED. Optical flux is the average flux of the data obtained from Boltwood
observatory optical telescope, KVA telescope, and WIYN telescope during the campaign period.

State\,4 and state\,9 \citep{flaring_state49} observations were taken from a comparatively longer time campaign
in 2003 and 2004. The X-ray flux obtained from {\it R}XTE were grouped into low-, medium- and high-flux groups.
For each X-ray observation in a given group, {\it Whipple} VHE\,$\gamma$-ray data that had been observed within
an hour of the X-ray data was selected. State\,4 (i.e., medium-flux) was observed between March 8 and May 3, 2003;
whereas state\,9 (i.e., high-flux) was observed on April 16-20, 2004. Optical datasets obtained with {\it Whipple}
Observatory's 1.2\,m telescope and Boltwood Observatory's 0.4\,m telescope were also selected based on the same
grouping method. The optical data measured during the whole campaign were not simultaneous with the other
multi-wavelength data: however, the optical flux was not found to vary significantly during the campaign, so
its highest and lowest values are taken to be reliable proxies of the actual values.

State\,5 and state\,7 \citep{Foss+08} data were taken on March 18-25, 2001 during a multi-wavelength campaign.
State\,7 denotes the peak of the March 19 flare, whereas state\,5 denotes a post-flare state on March 22 and 23. In
both cases the X-ray and VHE\,$\gamma$-ray data were obtained with, respectively, {\it R}XTE and the {\it Whipple}
telescope. The lowest and highest optical fluxes obtained during the whole campaign with the 1.2\,m Harvard-Smithsonian
telescope on Mt.\,Hopkins were used in the SEDs for states\,5 and 7, respectively.

State\,6 \citep{flaring_state126} observations were taken, using the same optical and X-ray instruments as
states\,1 and 2, during a decaying phase of an outburst in May 2008. The VHE\,$\gamma$-ray data were taken with
VERITAS. There are $\sim$2.5 hours of strictly simultaneous data.

State\,8 \citep{flaring_state8} data were taken during a multi-wavelength campaign on June 6, 2008: VERITAS,
{\it R}XTE and Swift/BAT, and WEBT provided the VHE\,$\gamma$-ray, X-ray, and optical data, respectively.

}

\section{Results and discussion}

To each of the datasets we apply our $\chi^2$-minimization procedure (see Fig.~\ref{fig:flowchart}).
The best-fit SSC models are plotted alongside the SED data in Fig.~\ref{fig:bestfit00}.

Obtaining truly best-fit SSC models of simultaneous blazar SEDs is crucial to measure the
SSC parameters describing the emitting region. The obtained optimal model is proven to be
unique because the SSC manifold is thoroughly searched for the absolute $\chi^2$ minimum.
Furthermore, the very nature of our procedure ensures that there is no obvious bias affecting
the resulting best-fit SSC parameters.

{
Nevertheless, as it has also been discussed \citep{arXiv:10123754}, there may be caveats
related with the $\chi^2_\nu$ fitting, especially when applied to non-linear models such
as the present one. For this reason, it is important to try to understand the goodness
of the fit with methods other than the value of $\chi^2_\nu$ (reported in table~\ref{tab:results01}).
Following \citep{arXiv:10123754}, we have applied the Kolmogorov-Smirnov (KS) test for
normality of the residuals of all our SED fits. A standard application of the test shows
that in all cases the residuals are \emph{not} normally distributed: the KS test thus fails
at the 5\% significance level. It is, of course, crucial to understand the reason for this
behavior of our SSC fits with respect to the KS test. Let us start with some
general remarks about the modelling of blazar emission and their observations. First, the one
zone SSC model contains two distinct physical processes in one same region, i.e. synchrotron
emission and its Compton up-scattered counterpart, that manifest themselves as essentially
separate components at very different energies; on the other hand, additional subtle effects
may enter the modelling of blazar emission, so that the SSC model may only be an approximation
to the real thing. (A more refined KS analysis suggests exactly this, see below). Second, our
blazar datasets do cover the (far apart) energy ranges spanned by, respectively, the synchrotron
and Compton emission: but they markedly differ in these two spectral regions, in that the
uncertainties associated with VHE data are much larger than those associated with the optical
and X-ray data.

Both these observations suggest us and give us the possibility of a slightly different approach
to the problem of the statistical significance of the fits, which will turn out to be quite
enlightening. We will refer to this other approach as the \emph{piecewise KS test}: it consists
in applying the KS test, separately to low energy and high energy data (see Appendix). The main
motivation behind this idea is, as discussed above, the marked difference, from both the physical
and the data-quality point of view, of the lower and higher energy ranges. Surprisingly, if for
each SED we separately check the low- and high-energy residuals for normality, the KS test always
confirms their (separate) normality at the 5\% confidence level. On the technical side we took all
the necessary precautions to improve reliability: the null hypothesis statistics have been obtained
from a Montecarlo simulation of 100,000 datasets having the same dimension as the residuals datasets.
Critical values to test normality of the residuals have been obtained from these numerically
determined statistics and, normality holds in all that cases we have
considered. Hence, it is unlikely to be a coincidence, and this calls for an explanation.
Clearly, the fact that the piecewise KS test is not able
to reject the normality of the low-/high-energy residuals separately, means that the
fitted SEDs can be considered as a reasonable models separately at low and high energies.
Of course, because of the quality of especially high energy data, the uncertainties on the
parameters of the fit are sometimes quite large, a fact that unfortunately can not be avoided,
despite the fact that the quality of the datasets that we are using is certainly above
average among those that are available. At the same time, the failure for normality of the
residuals on the standard KS test, in which low- and high-energy residuals are considered
simultaneously, suggests that existing data may require our adopted SSC model to be improved --
perhaps by taking into account higher-order effects (e.g., a multiple-break or curved electron
spectrum). So, and
with more and better VHE data available, it could be possible to reduce the uncertainties in
the parameters and to obtain a higher overall significance in the fit.

Given the above, the results found with our fits are the best possible in the framework of
existing datasets and models of blazar emission: at least, although preliminary, they have
a quantitative and clear statistical meaning. Therefore, we think it is useful to examine
some of their possible consequences. We'll do so in what follows.
}

\begin{figure*}
\centerline{\includegraphics[width=16.5cm]{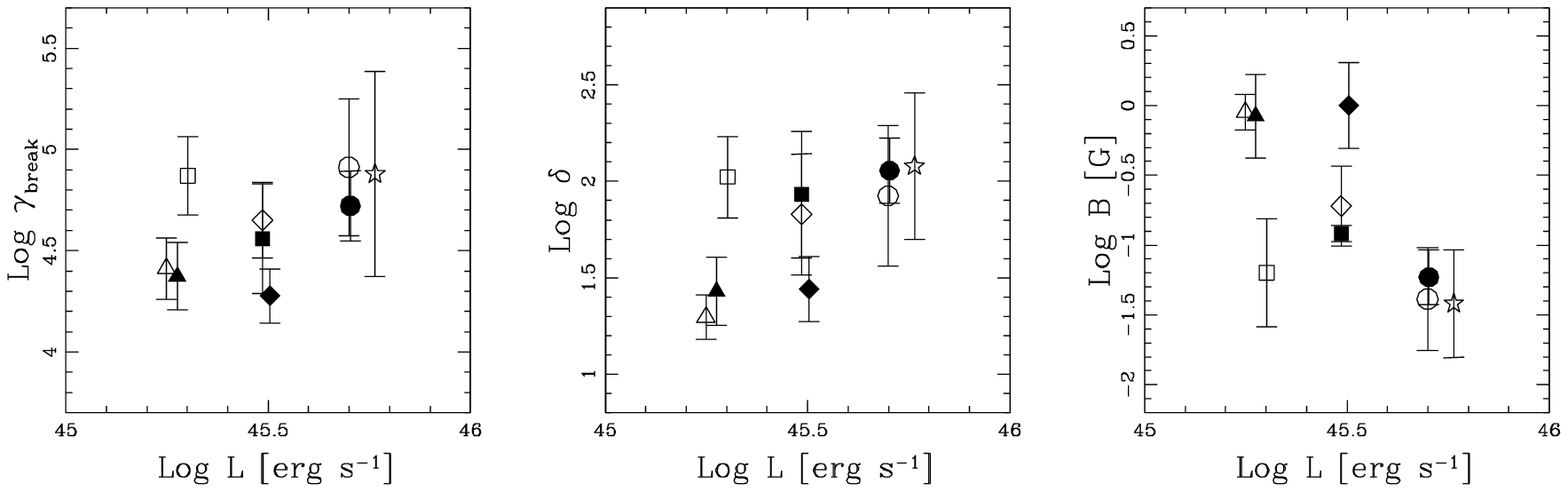}}
\vspace*{4mm}
\centerline{\includegraphics[width=16.5cm]{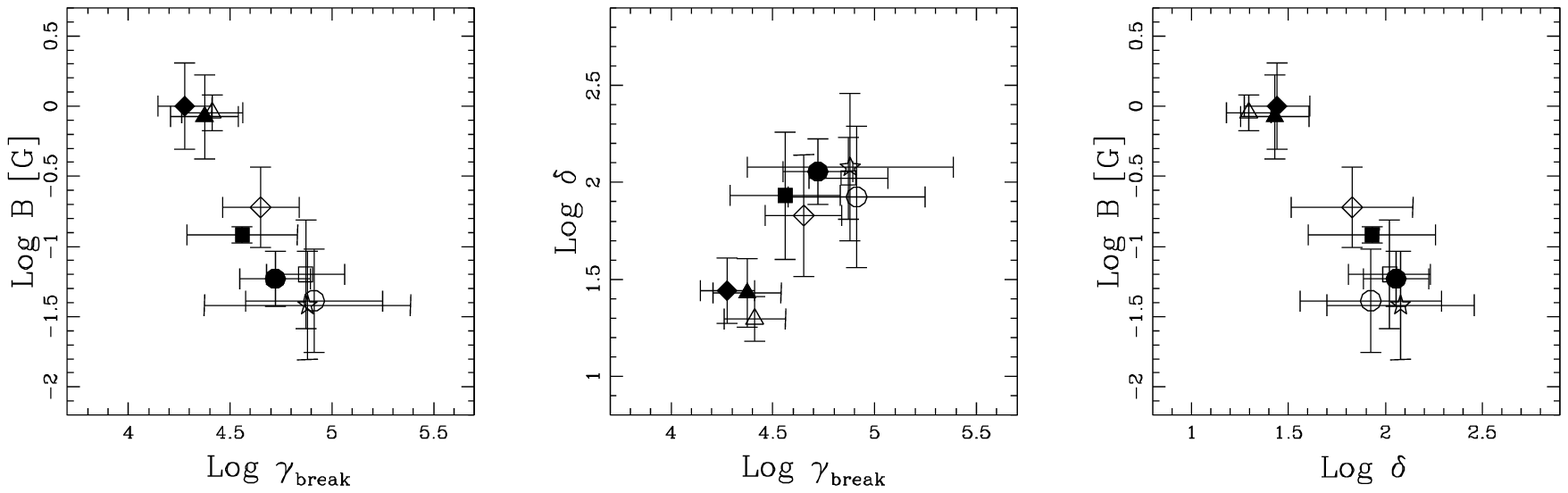}}
\begin{center}
\caption{\label{fig:bestfit01}{\it Top.} Variations of the SSC parameters magnetic field $B$, Lorentz
factor $\delta$, and $\gamma_{\rm br}$ as a function of the model's bolometric
luminosity. The other SSC parameters that are left to vary ($R$, $n_e$, $\gamma_2$,
$n_1$, $n_2$ only show a scatter plot with the luminosity.
{\it Bottom.} Correlations between $B$, $\delta$, and $\gamma_{\rm br}$.}
\end{center}
\end{figure*}

In Tables~\ref{tab:results01}, \ref{tab:results02} we report the best-fit SSC parameters. Source activity
(measured as the total luminosity of the best-fit SSC model) appears to be correlated with $B$, $\gamma_{\rm br}$,
and $\delta$ (see Fig.~\ref{fig:bestfit01}-{\it top}) -- and to be uncorrelated with the remaining SSC parameters.
The bolometric luminosity used in these plots has been obtained directly from the fitted SED. In more detail,
after determining the numerical approximation to the SED $\log [\nu F(\nu)]$, the parameters being fixed at
their best values obtained with the previously described minimization procedure, we have performed
$L = \int_{\nu _{\mathrm{min}}} ^{\nu_{\mathrm{max}}} \nu F (\nu) d \nu$,
with $\nu_{\mathrm{min}}$, $\nu_{\mathrm{max}}$ set at $2.5$ decades, respectively, below the synchrotron
peak and above the Compton peak. In this way we make sure to perform the integral over
all the relevant frequencies in a way that is independent from any in location of these peaks.

{
We then searched the data plotted in the top row of Fig.~\ref{fig:bestfit00} for possible correlations.
The linear-correlation coefficients turn out to be $0.67$, $0.64$ and $0.54$, which confirm linear
correlations with confidence levels of $4.8\%$, $6.3\%$ and $13.3\%$, respectively. As an additional
test (given the relatively low statistics of our datasets), we checked that the KS test confirms
normality of the fit residuals
\footnote{The same approach involving a Montecarlo generated empirical distribution
    for the null-hypothesis described just above has been used
    also in all these cases.
}.

{
All the parameters derived through our automatic fitting procedure are
within the range of SSC parameters found in the literature for HBLs in
general (e.g. \citep{Tav+01, Tav+10, Taglia+08, CelGhis08}) and for
Mrk\,421 in particular (e.g., \citep{BednaProt97, Tav+98, Mar+99, Ghis+02,
Kon+03, Foss+08}). In particular, large Doppler factors such as those
derived in our extreme cases, $\delta>50$, have been occasionally derived
(e.g., \citep{Kon+03, Foss+08}; see also the discussion in \citep{Ghis+05}). }

As is seen from Fig.~\ref{fig:bestfit01}-{\it top}, $\gamma_{\rm br}$ and $B$ are correlated, respectively, directly
({\it left}) and inversely ({\it right}) with $L$. This may be explained as follows. An
increase of $\gamma_{\rm br}$ implies an effective increase of the energy of most electrons
(or, equivalently, of the density of $\gamma<\gamma_{\rm br}$ electrons). To keep the
synchrotron power and peak roughly constant (within a factor of $3$; see Fig.~\ref{fig:bestfit00}), $B$ must decrease. This
improves the photon-electron scattering efficiency, and the Compton power increases. The
total (i.e., synchrotron plus Compton) luminosity will be higher. So a higher $\gamma_
{\rm br}$ implies a lower $B$ and a higher emission state. The $\delta$--$L$ correlation
(Fig.~\ref{fig:bestfit01}--{\it top-middle}) results from combining the $B$--$\delta$ inverse correlation
(Fig.~\ref{fig:bestfit01}--{\it bottom-right}; see below) and the $B$--$L$ anticorrelation.

A deeper insight on emission physics can be reached plotting the three $L$-dependent
parameters one versus the other (Fig.~\ref{fig:bestfit01}-{\it bottom}). The $B$--$\gamma_{\rm br}$ anticorrelation,
with $\Delta{\rm log \,B} \simeq -2\, \Delta {\rm log}\, \gamma_{\rm br}$
(Fig.~\ref{fig:bestfit01}--{\it bottom-left}), derives from the synchrotron peak, $\nu_{\rm s} \propto B \gamma^2$,
staying
roughly constant (see Fig.~\ref{fig:bestfit00}). For fixed synchrotron and Compton peak frequencies
in a relativistically beamed emission, the $B$--$\delta$ relation is predicted to be
inverse in the Thompson limit and direct in the Klein-Nishina limit \citep[e.g.,][]{Tav+98}:
because $\nu_{\rm s}$, $\nu_{\rm c}$ do not greatly vary from state to
state in our data (see Fig.~\ref{fig:bestfit00}), the correlation in Fig.~\ref{fig:bestfit01}--{\it bottom-right}
suggests that the
Compton emission of Mrk\,421 is always in the Thompson limit. The $\delta$--$\gamma_{\rm
br}$ correlation (Fig.~3--{\it bottom-middle}) results as a corollary of the condition of constant
$\nu_{\rm s}$, $\nu_{\rm c}$ emitted by a plasma in bulk relativistic motion toward the
observer.

{
Our fits show clear trends among some of the basic physical quantities
of the emitting region, the magnetic field, the electron Lorentz factor
at the spectral break, and the Doppler factor (see Fig.~\ref{fig:bestfit01}). In particular,
$B$ and $\gamma_{\rm break}$ follow a relation $B \propto \gamma_{\rm
break}^{-2}$, while $B$ and $\delta$ are approximately related by $B
\propto \delta^{-2}$.

Rather interestingly, the relation connecting $B$ and $\gamma_{\rm}$ is
naturally expected within the context of the simplest electron acceleration
scenarios (e.g., \citep{Henri+99}). In this framework, the typical
acceleration timescale, $t_{\rm acc}$, is proportional to the gyroradius:
$t_{\rm acc}(\gamma) \propto r_{\rm L}/c$, where $r_{\rm L}=\gamma m_{e}
c/(e B)$ is the Larmor radius. On the other hand, acceleration competes
with radiative (synchrotron and IC) cooling. In Mrk\,421, characterized by
comparable power in the synchrotron and IC components, we can assume that
$t_{\rm cool}(\gamma) \approx t{\rm syn}\propto 1/\gamma B$. The maximum
energy reached by the electrons is determined by setting these two timescales
equal, i.e.
\begin{equation}
t_{\rm acc}(\gamma _{\rm max})=t_{\rm cool}(\gamma _{\rm max}) \,
\rightarrow
\, \gamma_{\rm max} \propto B^{-1/2} \,,
\end{equation}
in agreement with the relation derived by our fit. It is therefore tempting
to associate $\gamma_{\rm break}$ to $\gamma_{\rm max}$ and explain the
$B\propto \gamma_{\rm break}^{-2}$ relation as resulting from the
acceleration/cooling competition.

If the above inference is correct, one can explain the variations of $\gamma_
{\rm break}$ as simply reflecting the variations of $B$ in the acceleration
region. In turn, the variations of $B$ could be associated either to changes
in the global quantities related to the jet flow or to some local process in
the jet (i.e. dissipation of magnetic energy through reconnection). The second
relation mentioned above, i.e. $\delta$ vs. $B$, seems to point to the former
possibility. First of all let us assume $\delta \sim \Gamma$ (since we are
probably observing the Mrk\,421 jet at a small angle w.r.t. the line of sight).
A general result of jet acceleration models is that, during the acceleration
phase, the jet has a parabolic shape, $R \propto d^{1/2}$ (with $d$ the distance
from the central black hole), and the bulk Lorentz factor $\Gamma$ increases with
$d$ as $\Gamma \propto d^{1/2}$ (e.g., \citep{VlaKon04, Kom+07}.
On the other hand, if the magnetic flux is conserved, $B\propto R^{-2}
\propto d^{-1}$ and thus we found $B\propto \Gamma ^{-2}$. Albeit
somewhat speculative, these arguments suggest that the trends displayed in Fig.~\ref{fig:bestfit01}
are naturally expected in the general framework of jet acceleration. }
}

The correlations in Fig.~\ref{fig:bestfit01}-{\it bottom} seem to be tighter than those in
Fig.~\ref{fig:bestfit01}-{\it top}. The larger
scatter affecting the latter owes to the fact that the electron density, $n_e$,
that also enters the definition of SSC luminosity, shows no correlation with
$B$, $\gamma_{\rm br}$, and $\delta$ -- hence it slightly blurs the latter's
plots with luminosity.

One further note concerns error bars. Our code returns $1$-$\sigma$ error bars.
To our best knowledge, this is the first time that formal errors of SED fits
are obtained in a rigorous way. As an example of the soundness of the method,
note that the obtained values of $\delta$ are affected by the largest errors
when the distribution of the VHE data points is most irregular (e.g. states 4, 7).

We notice that the variability of Mrk\,421 markedly differs from that of (e.g.)
the other nearby HBL source, Mrk\,501. The extremely bursting state of 1997 showed
a shift of $\nu_{\rm s}$ and $\nu_{\rm c}$ by two and one orders of magnitude,
respectively, suggesting a Klein-Nishina regime for the Compton peak \citep{Pian+98}.
Based on the data analyzed in this paper, Mrk\,421 displays (within our
observational memory) a completely different variability pattern. However, one
important similarity may hold between the two sources: based on eyeball-fit
analysis of Mrk\,501's SED in different emission states using a SSC model similar
to the one used here, \citet{Acc+10} suggest that $\gamma_{\rm br}$ does
vary with $L$. If this correlation is generally true in blazars, the implication
is that particle acceleration, providing fresh high-energy electrons within the blob,
must be one defining characteristic of excited source states.

\begin{table*}
\begin{center}
{\footnotesize%
\begin{tabular}{ l c c c | c}
\noalign{\smallskip}
\hline
\hline
\noalign{\smallskip}
Source & $B$ & $R$ & $\delta$ & $\chi _{\nu} ^{2}$ \\
       &[gauss]&[cm]&         & \\
\noalign{\smallskip}
\hline
\noalign{\smallskip}
%
State 1 & $(9 \pm 3) \times 10 ^{-1}$ & $(9 \pm 4) \times 10 ^{14}$ & $(2 \pm 0.5) \times 10 ^{1}$ & $0.84$\\
State 2 & $(8 \pm 6) \times 10 ^{-1}$ & $(8 \pm 4) \times 10 ^{14}$ & $(2.7 \pm 1.1) \times 10 ^{1}$ & $1.86$\\
State 3 & $(6 \pm 6) \times 10 ^{-2}$ & $(2.0 \pm 1.5) \times 10 ^{15}$ & $(1.0 \pm 0.5) \times 10 ^{2}$ & $0.91$\\
State 4 & $(1.21 \pm 0.16) \times 10 ^{-1}$ & $(1.1 \pm 1.3) \times 10 ^{15}$ & $(8 \pm 6) \times 10 ^{1}$ & $0.89$\\
State 5 & $(1.9 \pm 1.3) \times 10 ^{-1}$ & $(10 \pm 4) \times 10 ^{14}$ & $(7 \pm 5) \times 10 ^{1}$ & $0.67$\\
State 6 & $1.0 \pm 0.7$ & $(6 \pm 3) \times 10 ^{14}$ & $(2.8 \pm 1.1) \times 10 ^{1}$ & $1.39$\\
State 7 & $(4 \pm 3) \times 10 ^{-2}$ & $(2 \pm 5) \times 10 ^{15}$ & $(8 \pm 7) \times 10 ^{1}$ & $1.61$\\
State 8 & $(6 \pm 3) \times 10 ^{-2}$ & $(2 \pm 1.8) \times 10 ^{15}$ & $(1.1 \pm 0.4) \times 10 ^{2}$ & $0.60$\\
State 9 & $(4 \pm 3) \times 10 ^{-2}$ & $(2 \pm 4) \times 10 ^{15}$ & $(1.2 \pm 1.0) \times 10 ^{2}$ & $0.85$\\
\noalign{\smallskip}
\hline
\hline
\end{tabular}}
\end{center}
\caption{Best-fit single-zone SSC model parameters for the nine datasets of Mrk\,421.
States are named as in Fig.~\ref{fig:bestfit00}.}
\label{tab:results01}
\end{table*}
\begin{table*}
\begin{center}
{\footnotesize%
\begin{tabular}{ l c c c c c }
\noalign{\smallskip}
\hline
\hline
\noalign{\smallskip}
Source & $n_{\rm e}$ & $\gamma_{\rm br}$ & $\gamma_{\rm max}$ & n$_1$ & n$_2$ \\
       & [cm$^{-3}$] &                    &                   &                    &     \\
\noalign{\smallskip}
\hline
\noalign{\smallskip}
%
State 1 & $(1.3 \pm 1.5) \times 10 ^{3}$ & $(2.6 \pm 0.9) \times 10 ^{4}$ & $(1.05 \pm 0.18) \times 10 ^{7}$ & $1.49 \pm 0.19$ & $3.77 \pm 0.11$\\
State 2 & $(1 \pm 3) \times 10 ^{3}$ & $(2.4 \pm 0.9) \times 10 ^{4}$ & $(4.1 \pm 1.1) \times 10 ^{6}$ & $1.5 \pm 0.3$ & $3.62 \pm 0.14$\\
State 3 & $(5 \pm 5) \times 10 ^{3}$ & $(7 \pm 3) \times 10 ^{4}$ & $(7 \pm 5) \times 10 ^{7}$ & $2.05 \pm 0.10$ & $4.8 \pm 0.3$\\
State 4 & $(2 \pm 5) \times 10 ^{3}$ & $(4 \pm 2) \times 10 ^{4}$ & $(8.2 \pm 1.7) \times 10 ^{6}$ & $1.8 \pm 0.3$ & $4.11 \pm 0.13$\\
State 5 & $(2 \pm 5) \times 10 ^{3}$ & $(4.5 \pm 1.9) \times 10 ^{4}$ & $(2.4 \pm 0.3) \times 10 ^{7}$ & $1.7 \pm 0.3$ & $4.3 \pm 0.180$\\
State 6 & $(4 \pm 4) \times 10 ^{3}$ & $(1.9 \pm 0.6) \times 10 ^{4}$ & $(1.8 \pm 0.4) \times 10 ^{6}$ & $1.54 \pm 0.11$ & $4.37 \pm 0.09$\\
State 7 & $(1 \pm 7) \times 10 ^{3}$ & $(8 \pm 6) \times 10 ^{4}$ & $(7 \pm 2) \times 10 ^{6}$ & $1.7 \pm 0.4$ & $4.23 \pm 0.20$\\
State 8 & $(4 \pm 9) \times 10 ^{1}$ & $(5 \pm 2) \times 10 ^{4}$ & $(1.6 \pm 0.4) \times 10 ^{7}$ & $1.5 \pm 0.2$ & $4.22 \pm 0.14$\\
State 9 & $(1 \pm 7) \times 10 ^{2}$ & $(8 \pm 9) \times 10 ^{4}$ & $(1.1 \pm 0.4) \times 10 ^{7}$ & $1.6 \pm 0.5$ & $3.9 \pm 0.2$\\
\noalign{\smallskip}
\hline
\hline
\end{tabular}}
\end{center}
\caption{Best-fit single-zone SSC model parameters for the nine datasets of Mrk\,421.
Numbering convention as in Fig.~\ref{fig:bestfit00}.}
\label{tab:results02}
\end{table*}

\acknowledgements
We thank Daniel Gall for providing the datasets corresponding to states 1, 2 and 6, {and an anonymous
referee for useful comments and suggestions.} One of us (SA)
acknowledges partial support from the long-term Workshop on Gravity and Cosmology (GC2010: YITP-T-10-01)
at the Yukawa Institute, Kyoto University, during the early stages of this work, {and warmly thanks
P.\,Creminelli and S.\,Sonego for insightful discussions on some topics touched upon in this paper.}

\vskip 1.5truecm

\appendix

\centerline{\bf APPENDIX}
\centerline{\bf A TOY MODEL MIMICKING THE PIECEWISE KS TEST RESULTS FOR THE SED'S}

{In this appendix we reproduce the outcome of the piecewise KS approach to goodness of fit
using a simulated toy model. Let us consider the function $p(x ; a , b , c) := a x^4 + b x^2 + c$.
By using a pseudo-random generator, that produces uniformly distributed pseudo-random numbers in the
interval $[0,1]$ -- which we will henceforth call \texttt{rnd}  --, we generated $(x_i , y_i)_{i=1 ,
\dots{} , 5}$ pairs of points and $(w_i)_{i=1, \dots{} , 5}$ numbers, where $x_{i} = - 2.3 + 0.1 \cdot
( i + 0.1 \cdot \texttt{rnd} )$, $y_i = p ( -2.3 + 0.1 \cdot i ; -1 , 4 , 0) + 0.2 \cdot \texttt{rnd}$ and $w_i =
\sigma_i^{-2}$ with $\sigma_i = 0.01 \cdot ( \texttt{rnd} + 1 )$. These can be considered as $5$ measurements
around the $x = -2$ local maximum of the function $-x^4 + 4 x^2$ with small uncertainties $\sigma_i$.
We then generated another $5$ pairs of points $(x_i , y_i)_{i=6 , \dots{} , 10}$ and $(w_i)_{i=6, \dots{},
10}$ numbers with $x_i = 1.7 + 0.1 \cdot ( i + 0.1 \cdot \texttt{rnd} )$, $y_i = p ( 1.8 + 0.1 \cdot i ; -2 , 16 ,
-14) + 0.3 \cdot \texttt{rnd}$, $w_i = \sigma_i^{-2}$ and, finally, $\sigma_i = 0.1 \cdot (\texttt{rnd} + 1)$,
which again can be considered as $5$ slightly randomized measurement around a maximum, but now the
maximum of the function $-2 x^4 + 8 x^2 + 1$ and with much larger uncertainties that the first five
ones.

By minimizing $\chi^2$ we can fit the above set of $10$ randomly generated data points versus the
function $p (x ; a , b , c)$. After obtaining the desired fitted parameters we calculate the residuals
and, using the KS test, check if they are normally distributed. The KS test rejects normal distribution
of the entire set of residuals at the 5\% significance level. On the other hand, if we separately apply
the KS test to the first, i.e. $\{i = 1 , \dots{} , 5\}$, and the second, $\{ i = 6 , \dots{} , 10\}$, subset
of residuals, the test confirms that the residuals are normally distributed. This exactly replicates the
behavior we obtained in the SED fits.
}

\end{document}